\documentclass[letter]{IEEEtran}

\usepackage{balance}
\usepackage{cite}

 \usepackage[utf8]{inputenc}
     \usepackage{setspace}
\usepackage{amsmath,amssymb,amsfonts}
\usepackage[compact]{titlesec}         % you need this package
\titlespacing{\section}{0pt}{0pt}{0pt} % this reduces space between (sub)sections to 0pt, for example

\usepackage{tikz}
\usetikzlibrary{shapes.geometric, arrows}
\usepackage{amsmath}

\tikzstyle{startstop} = [rectangle, rounded corners, minimum width=3cm, minimum height=1cm, text centered, draw=black, fill=red!30]
\tikzstyle{process} = [rectangle, minimum width=3cm, minimum height=1cm, text centered, draw=black, fill=orange!30]
\tikzstyle{parameter} = [ellipse, minimum width=3cm, minimum height=1cm, text centered, draw=black, fill=yellow!30]
\tikzstyle{arrow} = [thick,->,>=stealth]

\usepackage{caption}
\usepackage{graphicx}
\usepackage{subcaption}
\usepackage{algorithm}
\usepackage[noend]{algpseudocode}
\usepackage{epstopdf} %converting to PDF
\usepackage{textcomp}
\usepackage{xcolor}
\newcommand{\RN}[1]{%
\textup{\uppercase\expandafter{\romannumeral#1}}
}

\begin{document}
%\title{Advanced OBP for Enhanced Routing and Load Balancing in High Throughput Satellites}
\title{A Scalable Architecture for Future Regenerative Satellite Payloads}
\IEEEoverridecommandlockouts  
% \author{Authors}
\author{\IEEEauthorblockN{Olfa Ben Yahia,~\IEEEmembership{Member,~IEEE,} Zineb Garroussi,~\IEEEmembership{Member,~IEEE,} Brunilde~Sansò,~\IEEEmembership{Senior Member,~IEEE,} Jean-François Frigon,~\IEEEmembership{Senior Member,~IEEE,} Stéphane Martel, Antoine~Lesage-Landry,~\IEEEmembership{Member,~IEEE,} and Gunes~Karabulut~Kurt,~\IEEEmembership{Senior Member,~IEEE}}

\thanks{O. Ben Yahia, Z. Garroussi, B. Sansò, J. Frigon, A. Lesage-Landry, and G. Karabulut-Kurt are with the Department of Electrical Engineering, Polytechnique Montréal, Montréal, QC H3T 1J4, Canada.
Emails:\{olfa.ben-yahia, zineb.garroussi, brunilde.sanso, j-f.frigon, antoine.lesage-landry, gunes.kurt\}@polymtl.ca}
\thanks{S. Martel is with Satellite Systems, MDA, 21025 Trans-Canada Hwy Sainte-Anne-de-Bellevue, Qc H9X 3R2. Email: Stephane.Martel@mda.space} 
\thanks{
This work is supported by MDA, CRIAQ, MITACS, and NSERC.}}

\maketitle

%\addtolength{\topmargin}{0.2in}
\begin{abstract}

This paper addresses the limitations of current satellite payload architectures, which are predominantly hardware-driven and lack the flexibility to adapt to increasing data demands and uneven traffic. To overcome these challenges, we present a novel architecture for future regenerative and programmable satellite payloads and utilize interconnected modem banks to promote higher scalability and flexibility. We formulate an optimization problem to efficiently manage traffic among these modem banks and balance the load. Additionally, we provide comparative numerical simulation results, considering end-to-end delay and packet loss analysis. The results illustrate that our proposed architecture maintains lower delays and packet loss even with higher traffic demands and smaller buffer sizes.
%significantly outperforms traditional models in terms of 
%operational efficiency and capability to handle high data rates. 
\end{abstract}


%****************************************%
\section{Introduction}
%****************************************%
%\lipsum[1-2] % Dummy text for illustration

%\begin{multicols}{2}
%\lipsum[1-2] % Dummy text for illustration
%\end{multicols}

\begin{figure*}[t]
  \centering
  \includegraphics[width=0.9\textwidth]{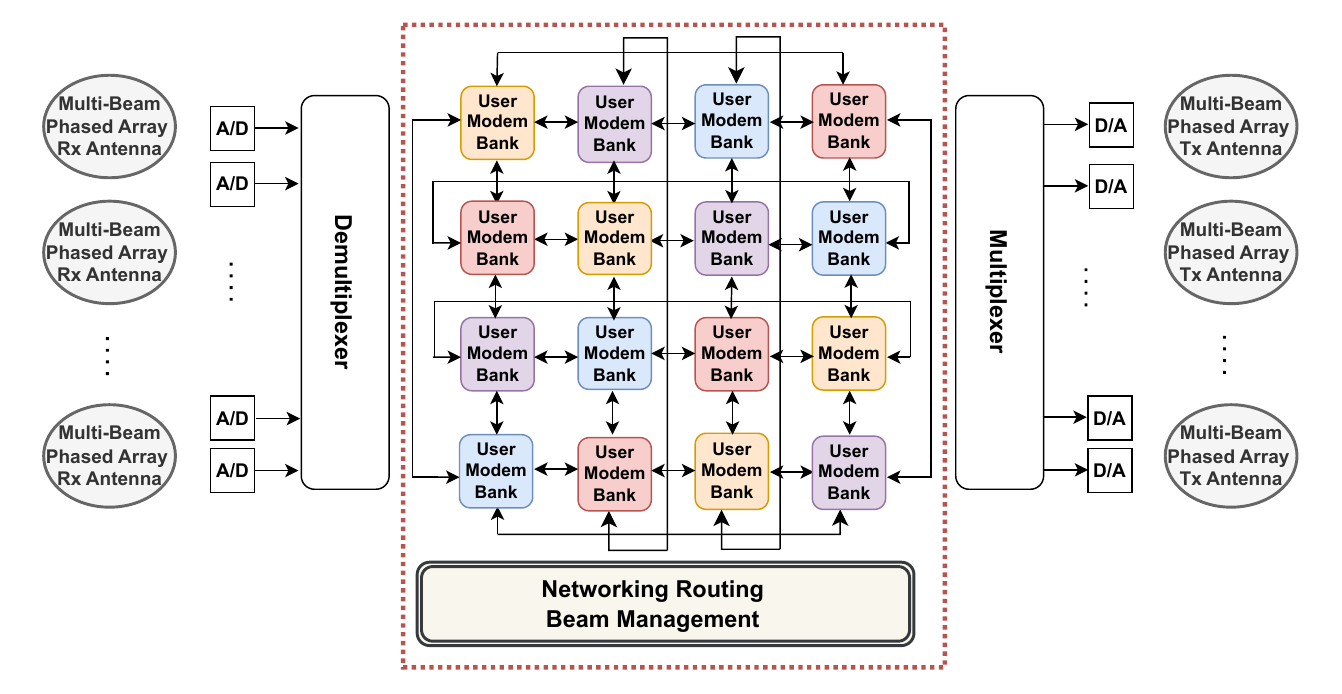}
  \caption{Overview structure of future regenerative scalable payload. Scalability is achieved through the use of multiple interconnected modem banks, which eliminates the need for a packet switch to connect them, thereby saving power, cost, and mass in the satellite system.}
  \label{figureOBP2}
\end{figure*}

Satellite technology has evolved significantly, enabling direct connectivity between satellites and standard mobile devices, often called direct-to-device or direct-to-cell (D2C) connectivity. This capability is a promising and crucial feature for future non-terrestrial networks, enabling satellite systems to reach billions of mobile users directly \cite{tuzi2023satellite}. In recent years, leading technology companies have been actively developing D2C connectivity by modifying their products to integrate with current satellite systems. For instance, Apple has significantly invested in enhancing Globalstar's satellite infrastructure, which includes 24 low Earth orbit (LEO) satellites and multiple ground stations. This improved network is crucial for supporting the Emergency SOS feature available on iPhone 14 models and beyond, providing users with vital communication capabilities in emergencies through satellites \cite{tuzi2023distributed}. Each element within a LEO constellation for D2C connectivity aims to mimic the service offered by a conventional terrestrial base station (BS) from space. This space-based BS could consist of either a single satellite or a cluster of smaller satellites \cite{tuzi2023distributed}.

Traditional terrestrial networks predominantly offer reliable connectivity in urban, suburban, and some rural areas \cite{manzoor2022improving}. However, the anticipated sixth-generation (6G) networks are expected to revolutionize this setup \cite{ITU-R-M2160-0}. In 6G, devices and machines will become the primary consumers of mobile data traffic. Consequently, one of the most challenging aspects of future network generations will be establishing satellite-to-device direct connectivity \cite{lee2023towards} and higher data rates of up to terabits-per-second. This capability will ensure comprehensive and seamless network access across diverse and remote locations, extending well beyond the current network infrastructure's limits.

A significant advancement in satellite technology is the transition from single-beam satellites to multi-beam satellites, also known as HTS. These satellites can provide higher data rates and serve more users \cite{itu-r2019key-elements}. The main features of HTS include increased data rates and wide coverage, multi-point beams for efficient spatial distribution, frequency multiplexing and beam hopping for dynamic bandwidth allocation, enhanced beam gain for better signal strength, advanced antennas for improved reception and transmission, and digital payload processing for more flexible and efficient data handling \cite{khammassi2023precoding}.

While significant advancements and progress have been made in developing devices for D2C connectivity, there has been no corresponding update in payload architecture to support the uneven traffic these devices will require. Current architecture primarily relies on a single packet switch component to connect the modem banks \cite{nguyen2020overview}. In this setup, incoming signals are converted into packets, and the payload functions as a packet switch, directing each packet to its intended destination. However, this packet switch component has major drawbacks: it consumes significant power and incurs substantial costs, both critical factors in satellite operations. Furthermore, the entire satellite operation is compromised if the packet switch fails, leading to potential mission failure. Additionally, with this existing architecture and available technologies, managing the anticipated massive traffic or achieving terabit-level data rates is nearly impossible. Moreover, this architecture is hardware-driven and lacks the flexibility and reconfigurability to adapt to traffic changes. This setup limits the ability to update or modify system operations in response to evolving traffic demand or operational requirements.

As satellite technologies evolve, innovative architectures are needed to handle this ever-growing demand. 
%Our paper proposes a novel architecture for satellite payloads designed to enhance connectivity and performance in response to this need. 
To address this need, in our work \cite{yahia2023evolution}, we present our vision for future high-throughput satellites (HTS), namely extremely-HTS. We discussed the related issues and technologies that could benefit such architectures. We propose an architecture based on multiple modem banks, which could be interconnected in various topologies. By leveraging layered software-defined networking (SDN) principles, this architecture aims to achieve several key objectives: efficient flow routing, effective load balancing among modem banks, and optimal resource allocation. This approach is intended to meet the requirements of end users by permitting real-time adaptations to changing operating conditions. However, our initial study in \cite{yahia2023evolution} does not delve into this architecture's mathematical details or specific use cases. Therefore, the present paper addresses this gap by proposing a mathematical formulation for this architecture. The following section details the proposed model that facilitates efficient resource management and decision-making processes within the satellite payload.

Unlike traditional payload designs based on packet switching, we propose a novel architecture for a flexible and programmable satellite payload to address the aforementioned issues, as illustrated in Figure \ref{figureOBP2}. This innovative design increases flexibility and adaptability, allowing the satellite system to respond to traffic changes and evolving data demands dynamically. The goal is to maximize throughput, minimize latency, and improve scalability compared to conventional designs.
Different from the current literature, the main contributions of this work can be summarized as follows:
\begin{itemize}
\item We explore the innovative architecture presented in \cite{yahia2023evolution}, which is based on interconnected multiple modem banks, and compare it with the current state-of-the-art centralized design. This architecture enhances satellite networks' flexibility and scalability.
\item We propose an optimization approach and validate it with simulations. This optimization generates routing tables, which are then used as input in the simulation process to enhance traffic management among modem banks.
\item We conduct numerical simulations to validate our proposed architecture, demonstrating that it maintains low delay and packet loss compared to traditional payload architectures with varying service rates.
\end{itemize}

The remainder of the paper is organized as follows: The system model and the optimization problem formulations are presented in Section~\ref{SEct2}. Numerical results and analysis are outlined
in Section~\ref{Sec4}. Finally, Section~\ref{Sec5} concludes the paper.

%****************************************%
\vspace{0.2cm}
\section{Proposed System Model}
\label{SEct2}
%****************************************
Recently, the Twins4Space project, for example, has been developing a new architecture of soft- and hardware components, including a distributed runtime environment where all nodes are connected by a SpaceWire network with a meshed topology that provides redundant data links for fault tolerance \cite{lindorfertwins4space}. This approach is gaining interest from international space agencies (e.g., ESA, NASA, and JAXA), who are investigating wireless technology to replace, complement, or extend wired data communication systems onboard satellites, aiming to improve flexibility and adaptivity of subsystem layouts and provide extra space inside the satellite \cite{buta2023wireless}.
Therefore, to enhance flexibility, maximize capacity and coverage, and achieve D2C connectivity with high data rates, we propose an architecture for future regenerative and configurable payloads, as depicted in Figure \ref{figureOBP2}. This satellite communication payload includes a variety of antennas, multiple interconnected modem banks, and a reconfigurable packet routing and switching engine. All incoming and outgoing traffic is fully regenerated on the satellite, with comprehensive routing, switching, and repacking processes occurring across different beams. The modems are equipped with a broad spectrum of modulation and coding options. Each modem is capable of adaptive coding and modulation tailored to each user link, providing the quality of service (QoS) required by each flow. Each modem bank has a buffer and queue at the input to handle incoming data efficiently. Additionally, the networking and routing subsystem is designed to support multiple levels of QoS, ensuring compliance with service-level agreements and facilitating best-effort networking solutions \cite{hindin2019technical, yahia2023evolution}.

Each modem bank can cover one or more cells in this proposed architecture. We suggest implementing 16 nodes connected in a toroidal topology to enhance the connectivity among the nodes and improve the routing processes within the satellite. This setup allows more efficient data routing between modem banks to ensure data is directed to the appropriate beam. In addition, to achieve a terabit-per-second data rate, each modem bank is designed to handle thousands of packets simultaneously, and the inter-modem links are equipped with gigabit-per-second connections. We propose an optimization problem to manage packet routing between different modem banks effectively. This approach generates routing tables that promote efficient data flows within the satellite's payload. 

\begin{figure*}[htpb]
    \centering
    \begin{subfigure}[t]{0.48\textwidth}
        \centering
        \includegraphics[width=\textwidth, height=0.25\textheight]{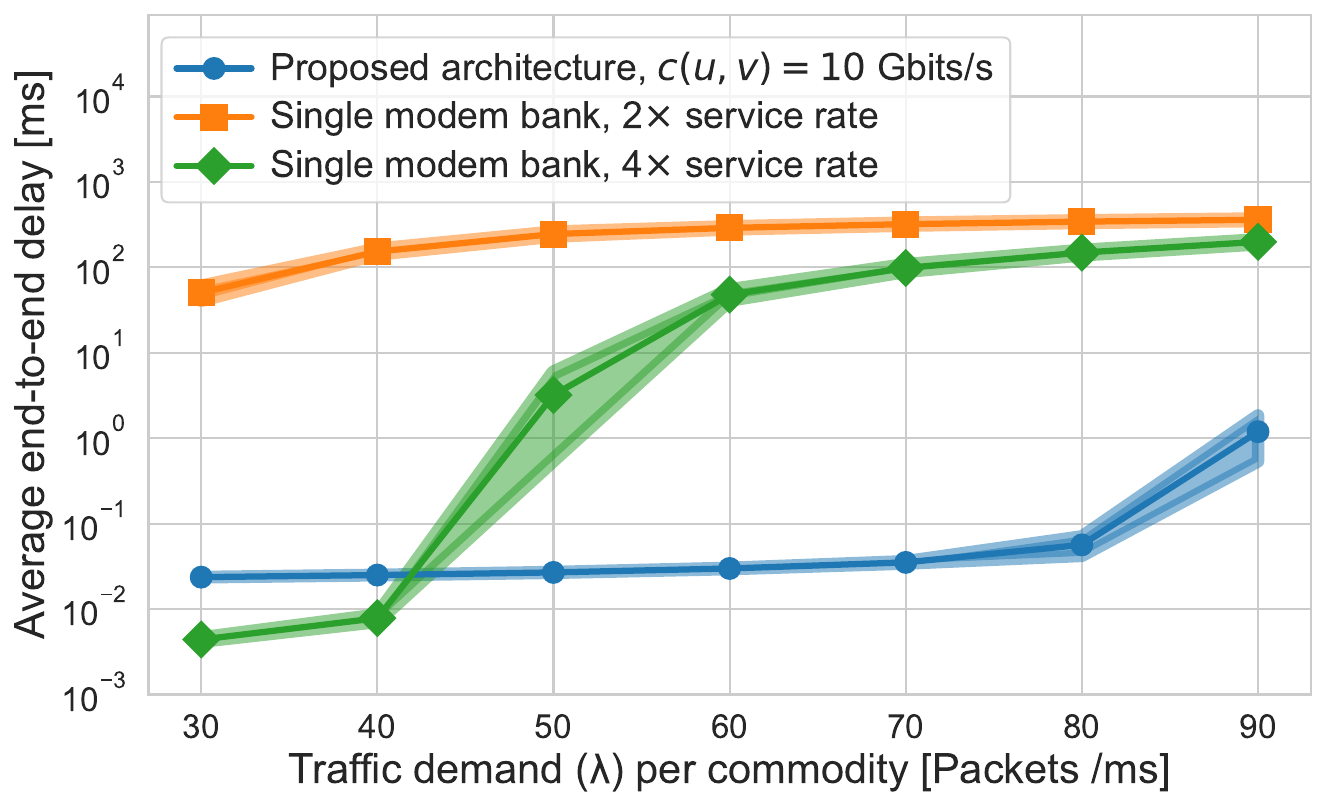}
        \caption{Average end-to-end delay.}
        \label{fig:delay6}
    \end{subfigure}
    \hfill
    \begin{subfigure}[t]{0.48\textwidth}
        \centering
        \includegraphics[width=\textwidth, height=0.25\textheight]{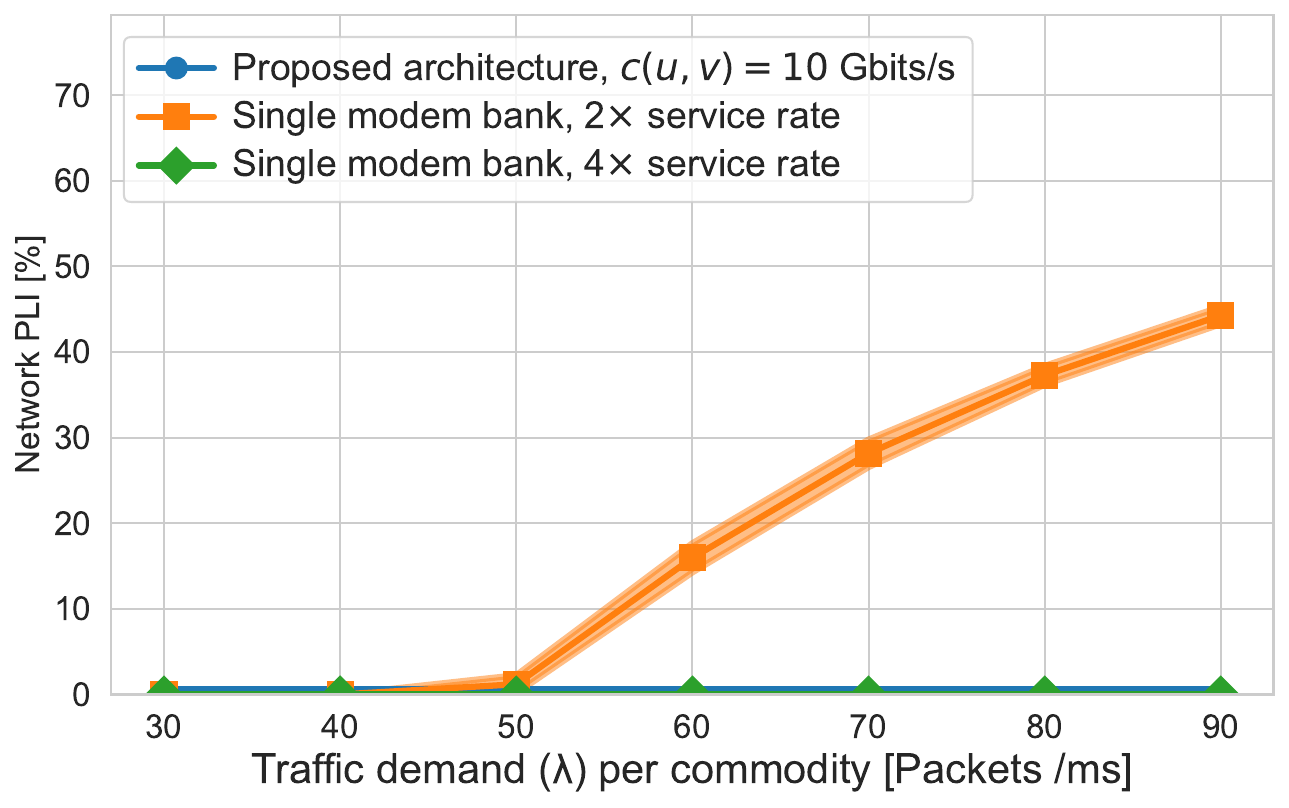}
        \caption{Packet loss indicator.}
        \label{fig:PLI6}
    \end{subfigure}
    \caption{Performance metrics for the network with a buffer size of $10^6$ packets.}
    \label{fig:network_metrics_106}
\end{figure*}
We model the satellite payload as a directed graph represented by \(\mathcal{G} = (\mathcal{V}, \mathcal{E})\). The different parameters used in this model are defined as follows: The set \{\(\mathcal{V}\)\} $\in \mathbb{N}$ represents the modem bank nodes. The set of directed edges \{\(\mathcal{E}\)\} indicates the connections or links between modem banks. \{\(\mathcal{K}\)\} represents the set of commodities, each defined by a tuple \( (s^k, t^k, d^k)\), specifying the source, destination, and required data flow, respectively. The set of paths \{\(\mathcal{P}^k\)\} for each commodity \(k \in \mathcal{K}\) run from the source \(s^k\) to the destination \(t^k\), with a maximum hop count to constrain latency. Paths through a specific link \((u,v)\)$\in \mathcal{E}$ are denoted as \(\mathcal{P}^k_{(u,v)}\).
The data rate of each link \((u, v)\) is noted as \(c(u,v) > 0\), indicating the maximum flow it can handle. The flow of a commodity \(k\) along a path \(p \in \mathcal{P}^k\) is represented as \(x_p^k\) and quantifies the amount of data transmitted. With this model, the optimization problem can be formulated as: \begin{align}
    \max_{x_p^k \in \mathbb{R}} \quad &  \min_{(u,v)\in \mathcal{E} }\left( c(u,v) - \sum_{k \in \mathcal{K}} \sum_{p \in \mathcal{P}^k_{(u,v)}} x_p^k \right) \label{Opt} \\
    \text{subject to} \quad & \sum_{p \in \mathcal{P}^k} x_p^k = d^k, \quad \forall k \in \mathcal{K} \label{C1} \\
    & \sum_{k \in \mathcal{K}} \sum_{p \in \mathcal{P}^k_{(u,v)}} x_p^k \leq c(u, v), \quad \forall (u, v) \in \mathcal{E} \label{C2} \\
    & x_p^k \geq 0, \quad \forall k \in \mathcal{K}, p \in \mathcal{P}^k. \label{C3}
\end{align}

In the above problem, the objective function eqn. (\ref{Opt}) aims to maximize the minimum residual capacity of the inter-modem links, ensuring an efficient and balanced distribution of data flows across the modem banks. Eqn. (\ref{C1}) represents the demand satisfaction constraint, which ensures that the total flow for each commodity \( k \) over all its paths equals the commodity's demand \( d^k \). This guarantees that each commodity's demand is fully met. The capacity constraint is given in eqn. (\ref{C2}), and it ensures that the total flow routed over any inter-modem link \( (u, v) \) does not exceed its maximum data rate \( c(u, v) \) to guarantee system stability.
Finally, the non-negativity constraint that ensures that all flow variables \( x_p^k \) are non-negative is presented in eqn. (\ref{C3}). 

The problem involves optimizing data flows within the payload, where multiple modem banks are interconnected by links with specific capacity \(c(u,v)\). Each link's capacity denotes the maximum data rate it can handle. The objective is to ensure data is transmitted efficiently from source to destination nodes without surpassing these capacities. The challenge is maximizing the links' minimum residual capacity and maintaining network efficiency.
%Furthermore, minimizing the number of modem transitions (hops) is crucial to reducing latency while ensuring that the data flow demand for each commodity is fully satisfied within these constraints.

\begin{figure*}[htbp]
    \centering
    \begin{subfigure}[t]{0.48\textwidth}
        \includegraphics[width=\textwidth, height=0.25\textheight]{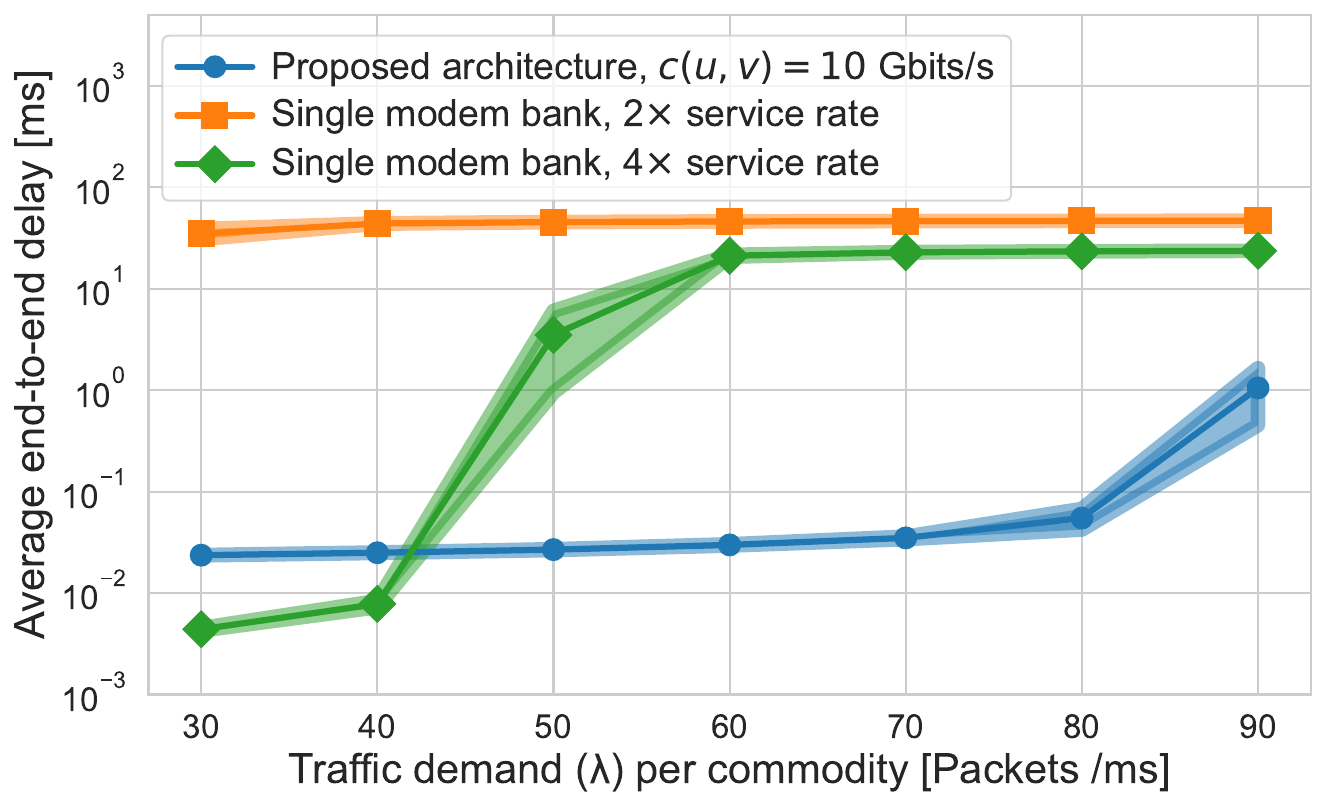}
        \caption{Average end-to-end delay.}
        \label{fig:delay4}
    \end{subfigure}
    \hfill
    \begin{subfigure}[t]{0.48\textwidth}
        \includegraphics[width=\textwidth, height=0.25\textheight]{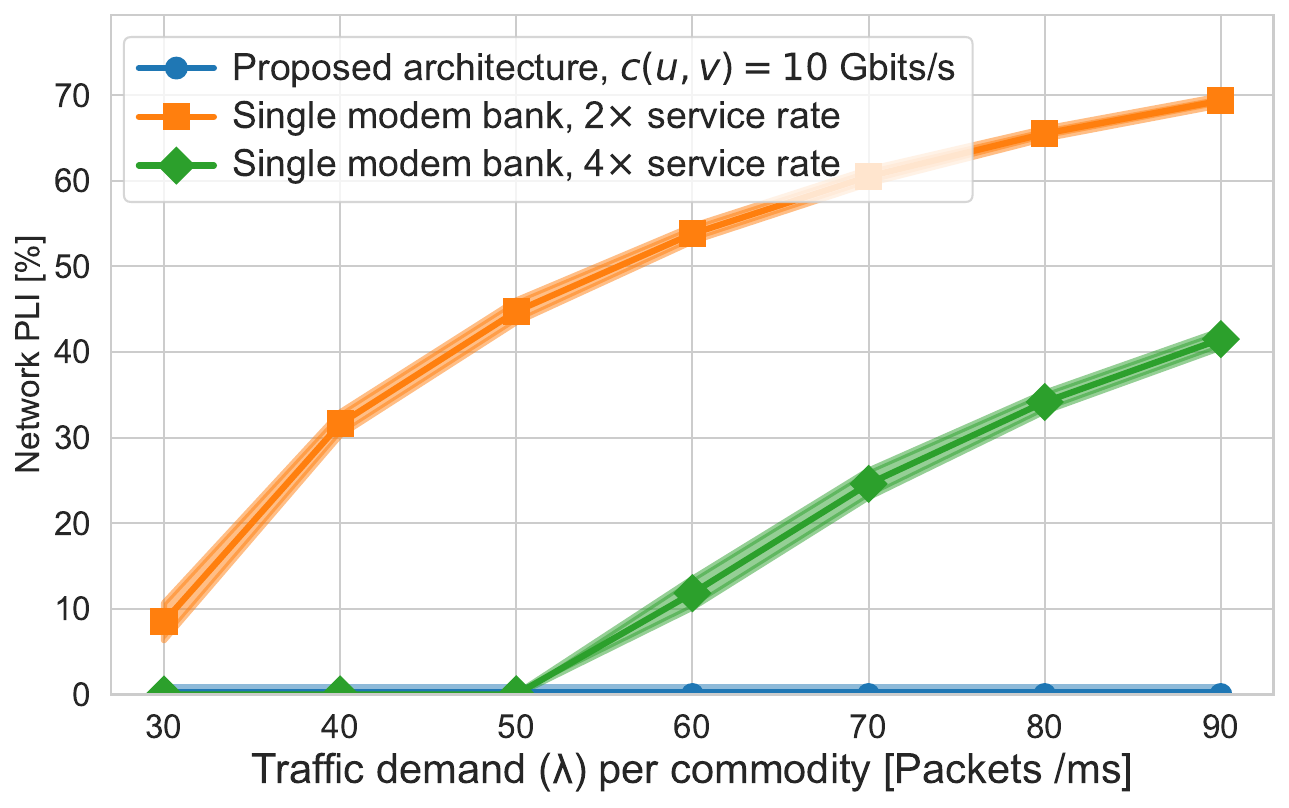}
        \caption{Packet loss indicator.}
        \label{fig:PLI4}
    \end{subfigure}
    \caption{Performance metrics for the network with a buffer size of $10^4$ packets.}
    \label{fig:network_metrics_104}
\end{figure*}

In the proposed simulation-based optimization framework, the flow routing process begins with inputting the incoming data flows into the optimization problem, as described by equations (\ref{Opt}) to (\ref{C3}). The optimization problem's solution includes the routing tables, which are then used as input for the simulation model. The simulation model evaluates the performance of the routing tables under various conditions, and the resulting performance metrics are analyzed to assess the effectiveness and efficiency of the proposed optimization approach.
%****************************************%
\vspace{0.2cm}
\section{Numerical Results and Analysis}
\label{Sec4}
%****************************************%

This section compares our proposed multi-modem bank architecture with a traditional single modem bank node, considering various processing rates as multiples of the service rate of a single modem bank in our proposed architecture. In this centralized architecture, routing tables are unnecessary as there is only one possible path, eliminating the need for optimization. Python simulations, combined with CVXPY \cite{diamond2016cvxpy} for optimization modelling using the GLPK solver \cite{GLPK}, are employed to assess performance metrics, including average end-to-end delay and packet loss indicator (PLI). Performance metrics are averaged, and $95\%$ confidence intervals are included. Due to computational resource limitations, we consider low values of processing rates. However, in real-world scenarios, the processing rate is expected to be higher to achieve terabits per second (Tbps) data rate and negligible delays. Table~\ref{Table_sim} illustrates the simulation parameters.
%This comparative analysis highlights the enhancements in throughput and reliability offered by our architecture.

\begin{table}[t]
\renewcommand{\arraystretch}{1.1}
    \caption{Simulation Parameters}
    \label{Table_sim}
    \centering
    \begin{tabular}{|p{4cm}|p{4cm}|}
        \hline
        \textbf{Parameter} & \textbf{Value} \\
        \hline
        Average packet size & 1500 Bytes \\
        Number of commodities $k$ & 8 \\
        Inter-modem bank link data rate $c(u,v)$ & $10$, $1$ Gbits/s \\
        Modem service rate & 100000 packets/s \\
        Buffer size & $10^4$, $10^6$ packets \\
      Arrival rate $\lambda$ & $30$k, $40$k, $50$k, $60$k, $70$k, $80$k, $90$k packets/s \\
	\hline
	\end{tabular}
\end{table}

Figure \ref{fig:network_metrics_106} shows the performance metrics with a buffer size of \(10^6\) packets. Figure \ref{fig:delay6} depicts the average end-to-end delay as a function of traffic demand per commodity. The proposed architecture with an inter-modem bank link capacity of \(c(u, v) = 10\) Gbits/s demonstrates consistently low delay across the increasing traffic demands. In contrast, single modem bank configurations with 2$\times$ and 4$\times$ service rates exhibit higher delays, particularly noticeable at higher traffic demands. Figure \ref{fig:PLI6} illustrates the PLI percentage. The proposed architecture maintains a negligible PLI. In contrast, both single modem bank configurations show a significant rise in PLI, with the 2$\times$ service rate configuration experiencing the highest packet loss as traffic demand increases. These results highlight the efficacy of the proposed architecture in maintaining low delay and packet loss under varying traffic conditions compared to traditional single modem bank configurations, even with high processing rates.

Figure \ref{fig:network_metrics_104} presents performance metrics for a network with a reduced buffer size of \(10^4\)packets. Figure \ref{fig:delay4} shows that the proposed architecture maintains consistently low delays despite the reduced buffer size. In contrast, the single modem bank configurations (2$\times$ and 4$\times$ service rates) show decreased delays compared to the buffer size of \(10^6\) packets. This is due to the reduced number of packets sent in the system. However, as shown in Figure \ref{fig:PLI4}, 2$\times$ and 4$\times$ service rate configurations exhibit significantly higher packet loss, especially at higher traffic demands. This illustrates that while the proposed architecture maintains stability with low packet loss, the traditional modem bank setups experience increased packet loss as buffer size decreases.

\begin{figure}[htbp]
  \centering
  \includegraphics[width=0.48\textwidth, height=0.25\textheight]{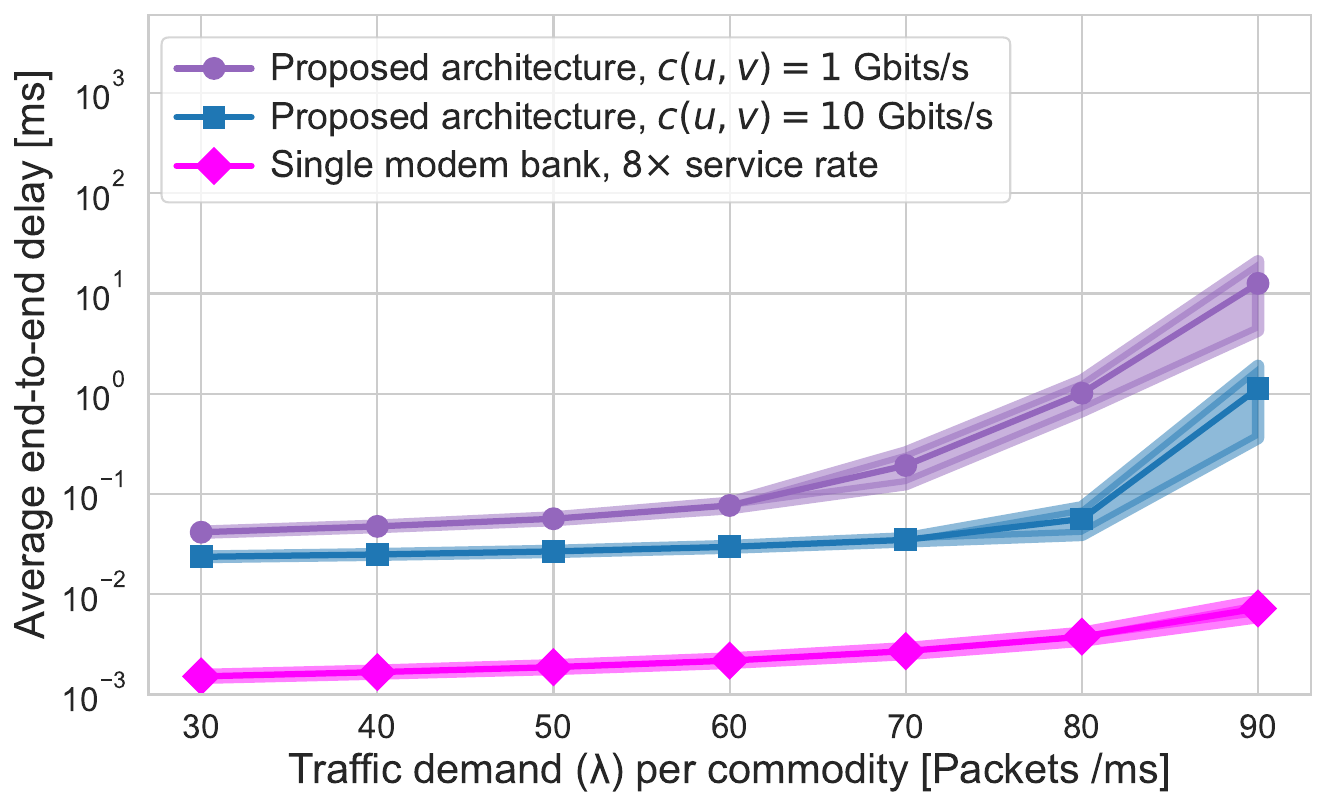}
  \caption{Performance metrics for the network with a buffer size of $10^4$ packets.}
  \label{fig:delay}
\end{figure}

Figure \ref{fig:delay} illustrates the average end-to-end delay for our architecture with varying inter-modem link rates and compares it with a single modem bank with an 8$\times$ service rate. The results show that increasing the inter-modem link capacity significantly improves the performance of our proposed architecture. Specifically, the configuration with an 8$\times$ service rate performs best, achieving the lowest delays. Therefore, to achieve the best performance metrics with the data set used, an 8$\times$ processing rate is recommended. However, it is important to note that implementing an 8$\times$ service rate solution is currently not feasible with the available technologies. While the 8$\times$ service rate solution is superior, the number of modems that can be integrated into a single modem bank using existing technology is highly limited, which restricts the scalability of such an architecture.

\section{Conclusion }
\label{Sec5}
%****************************************%
In this letter, we investigate a novel architecture for future programmable regenerative payloads. Based on multiple modem banks interconnected in a toroidal topology, this architecture utilizes an optimization problem for routing within these modems. Using the generated routing tables, we conducted simulations using Python. Considering various service rates, we compared our proposed architecture with a centralized single modem payload. The results showed that our proposed architecture maintains lower delays and packet loss, even for higher traffic demands and smaller buffer sizes.

While a configuration with an 8$\times$ processing rate could achieve better results, the number of modems that can be integrated into a single modem bank with existing technology is highly limited, restricting the scalability of such an architecture. Additionally, our proposed architecture offers greater scalability and flexibility, particularly in the face of node failures, where another node can take over, ensuring that the satellite mission is not compromised and continues functioning. This architecture also allows for increased regional coverage, making it a promising solution for future satellite networks.

\balance
\bibliographystyle{IEEEtran}
\bibliography{reference}
\end{document}